\documentstyle[aas2pp4]{article}

\tighten             

\newcommand{\hii}{H\,{\sc ii}\  }
\newcommand{\nii}{[N\,{\sc ii}]}
\newcommand{\oiii}{[O\,{\sc iii}]}
\newcommand{\ha}{H$\alpha$}
\newcommand{\hb}{H$\beta$}
\newcommand{\ohb}{${\rm{I[O\,II}] + \rm{I[O\,III}]}\over{\rm{I(H}\beta)}$}
\newcommand{\msun}{$M_{\odot}$}
\newcommand{\zsun}{$Z_\odot$}
\newcommand{\philb}{$\phi\over L_B$}
\newcommand{\ngc}{NGC\,}
\newcommand{\etal}{{\it et al.}\rm }
\newcommand{\etc}{{\it etc.}\rm }
\newcommand{\eg}{{\it e.g.}\rm }
\newcommand{\tauBl}{\tau_{\rm B}^{\rm l}}
\newcommand{\tauBc}{\tau_{\rm B}^{\rm c}}
\newcommand{\IZw}{I\,Zw\,18}

\begin{document}

\title {EMBEDDED CLUSTERS IN GIANT EXTRAGALACTIC \hii REGIONS: \\ 
 III. EXTINCTION AND STAR FORMATION } 
 
\author {Y.\,D.~Mayya\altaffilmark{1,2} and T.\,P.~Prabhu\altaffilmark{2}}
\affil{$^1$Infra Red Astronomy Group, 
Tata Institute of Fundamental Research, Homi Bhabha Road, Colaba, 
Bombay 400 005 INDIA} 
\affil{$^2$Indian Institute of Astrophysics, Bangalore 560 034 INDIA } 

\affil{Electronic mail: ydm@tifrvax.tifr.res.in and tpp@iiap.ernet.in} 
\affil{To appear in Astronomical Journal} 

\begin{abstract}

A study of star formation is carried out on 35 giant extragalactic \hii
regions (GEHRs) in seven galaxies using optical photometric data in $BVR$ broad
bands and in the emission line of \ha. Interstellar extinction,
metallicity and nebular contributions to the broad bands are estimated
using spectroscopic data on these objects. Dimensionless diagrams
involving $B-V$ and $V-R$ colors and the flux ratio of Balmer line to $B$ band
continuum are used to study star formation. The cluster colors
indicate reduced extinction towards stellar continuum compared to the values 
derived from Balmer lines for the ionized gas. The frequency of detection of 
classical young ($t<3$~Myr) regions with only one burst of star formation 
is found to be low as compared to young regions with an accompanying
population rich in red supergiants from a previous burst ($t\sim$10~Myr).
Reduced extinction towards cluster stars, destruction of ionizing photons
and the existence of older population, often spatially unresolvable from the 
younger population, all conspire to make the observed Balmer line equivalent 
widths low in a majority of the GEHRs. A scenario of star formation is suggested 
which explains many of the observed properties of GEHRs, including the core-halo 
structure, reduced extinction for the radiation from stars as compared to that 
from the nebular gas, non-detection of young single burst regions and
the co-existence of two populations of different ages. 

\end{abstract}

\twocolumn		

\section{Introduction}     

\noindent
Giant extragalactic \hii regions (GEHRs) are sites of current star formation 
activity in galactic disks. A detailed investigation of star formation in
these relatively simpler systems is important towards understanding star
formation in more complex star-forming systems such as blue compact 
galaxies (BCDs or \hii galaxies), and starburst systems. GEHRs in an 
individual galaxy are known to differ from one another in many of the 
observational properties, but still have several properties in common. The 
common properties such as the constancy of luminosity and diameter of the 
brightest (also largest) \hii regions have been used earlier for estimating 
the distances to galaxies (Sandage \& Tammann 1974, Kennicutt 1979).
It is important to investigate the reasons for similarities as well as
differences between different \hii regions in order to make use of GEHRs as 
reliable distance indicators or tracers of star formation rates in galaxies. 

Initial Mass Function (IMF), age, history, metallicity, extinction, mass or
luminosity are among the most important parameters determining the properties
of GEHRs. In recent years, there have been several attempts to investigate the
IMF parameters under a variety of conditions (see Scalo 1986). Yet, the IMF 
slope of $\alpha=2.35$ as obtained by Salpeter~(1955) remains the most
accepted value. Humphreys \& McElory (1984) have favoured a much steeper IMF 
for high mass stars in the Galaxy, while Parker \& Garmany (1993)
obtained an IMF slope close to Salpeter's value for
the stars in 30 Doradus, both studies being based on star counts. 
Investigations based on the integrated properties of GEHRs have so far
remained inconclusive on the exact value of IMF slope and the lower and upper
cutoff masses. There are some indications that IMF may be flatter or deficient
in low mass stars in active starburst regions (Scalo 1987).

In this study we aim to investigate the star formation histories of GEHRs,
taking into account the possible IMF differences from one region to another.
It is however important to take into account the effects of 
(i) interstellar extinction, (ii) metallicity and (iii) nebular contributions 
on the observable quantities, before these quantities can be compared with
the synthetic models. 
Interstellar extinction is conventionally derived using the observed
ratios of Balmer lines (see \eg\ Shields 1990).
Balmer decrements as well as metallicities and nebular contributions can be 
derived from spectroscopic data on individual regions. With the above problems 
in mind, we use the $BVR$\ha\ data of only those GEHRs from Mayya (1994)
for which additional spectroscopy is available 
in the literature. The data, after applying corrections for extinction 
and nebular contribution, are compared with the evolutionary population 
synthesis model to investigate the properties of star formation in the sample 
regions. Analyses were performed on dimensionless diagrams, thus eliminating 
the possible errors due to the uncertainties in the estimation of distance to 
the source and/or the mass of the star-forming region. 
We find that the extinction experienced by the embedded cluster stars is
lower compared to the values derived from the Balmer decrement. These results
are of special interest, especially after a similar conclusion has been reached
by the independent study of Calzetti \etal\ (1994) using ultraviolet (UV) and 
optical spectra of central regions of galaxies. Our study also reveals a 
slightly complex history of star formation in GEHRs.

In sec.~2, we describe the sample and the method followed in
extracting the pure stellar continuum from the observed $BVR$ data. In sec.~3, 
the properties of the embedded cluster are compared with population synthesis 
models. Our results are discussed in the context of recent results in sec. 4. 
Physical models for star formation in GEHRs and the distribution of cluster 
stars, gas and dust are proposed in sec.~5. Sec.~6 summarizes the main results 
of this study.

\section{Sources of Data and Preliminary Treatment}       

The GEHR data for this study are taken from the $BVR$\ha\ photometric catalogue
of Mayya (1994, Paper I henceforth). The catalogue contains CCD aperture 
photometric data on nearly 180 regions in nine nearby galaxies. In these 
studies, special care was taken in ensuring that the contamination by the 
background is minimum and hence the derived magnitudes are genuinely from the 
embedded cluster and the surrounding nebulosity. The usage
of apertures rather than slits, as well as measurement of $BVR$ continuum and
\ha\ emission line fluxes, centered on the same pixel, makes these
measurements ideally suited for the investigation of star formation properties.
Spectroscopic data are available for about one fourth of these regions 
($\sim$45). The sensitiveness of the background
values on the catalogued colors is indicated by a photometric quality index in
the catalogue. Regions having photometric quality index 4 (background
uncertainties leading to errors in excess of 0.3 mag in $B-V$ and $V-R$
colors: see Appendix in Paper~I for a discussion on background subtraction 
errors in \hii region photometry) are dropped to eliminate background
dominated regions from the sample. The estimated rms errors for the entire
sample in the catalogue are $<0.15$ mag in colors and $<0.05$ dex in 
the ratio of \hb\ to blue band luminosity (\philb). The regions selected for 
the study here are relatively brighter and hence the
errors are expected to be smaller. We estimate the errors to be a factor of
two less as compared to that for the entire sample. 
Finally, we are left with 5 regions in \ngc 1365, 3 in \ngc 2903, 14 in 
\ngc 2997, 2 in \ngc 3351, 3 in \ngc 4303 in addition to the giant \hii 
complex \ngc 2363 in \ngc 2366 and the starburst nucleus of \ngc 5253.
Measurements using bigger apertures, which include more than one emitting 
knot, are also included around six regions in \ngc 2997. 

The spectroscopic data involving \oiii\  and Balmer lines are used to 
estimate the abundance of oxygen, electron temperature ($T_e$) and visual 
extinction ($A_v$(Bal)) using a semi-empirical approach (McCall \etal\ 1985). 
All the observed
quantities have to be corrected for interstellar extinction before 
comparing with the models.  Total extinction towards extragalactic
\hii regions consists of three parts --- (i) extinction from the interstellar
medium (ISM) of our galaxy ($A_v$(gal)), (ii) extinction from the ISM of the 
parent galaxy outside the \hii region ($A_v$(ext)) and (iii) extinction due to 
the dust mixed with the gas within the \hii region ($A_v$(int)). $A_v$(gal) 
values for the program galaxies are given in Table~1 for each of the
program galaxies, as tabulated by de Vaucouleurs {\etal}\ (1991). Other 
quantities in the table are distances to the
parent galaxies, number of regions used in the present study and the sources 
of spectroscopic data and distances.
We denote the visual extinction as derived from the Balmer decrement by
$A_v$, in the remaining part of the text, though it
contains contribution from all the three components. We start our 
study using $A_v$ to deredden the cluster and nebular quantities,
assuming that the spectroscopically derived values are valid for both the 
cluster continuum and the nebular radiation integrated over the apertures.
In the later part of this section, we suggest alternative prescriptions for
extinction corrections that are necessary to explain the present data set.

The usage of $A_v$ to deredden quantities related to embedded clusters is
appropriate only if the amount of dust residing internal to \hii\ regions is 
negligible. The large spread in 
extinction values for regions within a single galaxy implies non-uniform 
distribution of dust within the parent galaxy, either within or outside the
\hii region. There have been studies investigating the relative contribution 
of the dust mixed with the ionized gas, known as `internal extinction', to the 
total extinction from the parent galaxy (Lequeux \etal\ 1981).
These studies show the internal extinction to be non-negligible at least for
some regions. The extinction values derived using the radio thermal continuum
is statistically known to be higher than that derived using the 
Balmer decrement (Israel \& Kennicutt 1980). Caplan \& Deharveng (1986) have 
studied the effect of different dust configurations, including the dust
internal to the \hii regions to explain the observed discrepancy between the
radio and optically derived extinction values. Internal reddening is found to
be significant for some regions in the Large Magellanic Cloud (LMC).
We find the extinction properties of embedded cluster stars to be different 
from that of the ionized gas and hence suggest an alternate prescription
for extinction correction in GEHRs. 

The \ha\  band fluxes have contribution also from \nii\ lines at 
$\lambda$~6548, 6583 \AA. The observed \ha$+$\nii\ fluxes are first corrected 
for interstellar extinction using $A_v$ and the galactic
extinction curve (Seaton 1979). \ha\  fluxes are computed from the observed 
\ha$+$\nii\ fluxes using the spectroscopically determined 
${{\rm{[N\,II]}}/{{\rm{H}}\alpha}}$
ratios. Lyman continuum luminosities are derived from extinction corrected
\ha\  luminosities using

\begin{equation}
{N_L\over{\rm{ph\,s}^{-1}}} 
                = 7.32\times 10^{11} {{\rm H}\alpha\over{\rm{erg~s}^{-1}}}
\end{equation}
which assumes an ionization bounded case B nebula (Osterbrock 1989). 
The expected \hb\  luminosity $(\phi)$ is computed from this using 

\begin{equation}
{\phi\over{\rm{erg~s}^{-1}}}  = 4.78\times 10^{-13} {N_L\over{\rm{ph\,s}^{-1}}.}
\end{equation}
The treatment of broad-band magnitudes is slightly different.
First, an estimation of the contribution from the nebula (line $+$ continuum) 
within each band is made based on the derived $T_e$ and extinction corrected 
\ha\  luminosity. The strengths of all bright lines relative to \hb\ line, are 
taken from McCall \etal\ (1985). Free-free, free-bound and 2-photon emission 
mechanisms are considered in computing the nebular continuum. 
Details of these computations are presented in Mayya (1995). 
For a majority of the regions, the nebular corrections amount to $<5$\% in $B$ 
and $V$ bands and 15--25\% in the $R$ band, with emission lines being the main 
contributors. The estimated nebular contributions to $B$ and $V$ bands exceed 
that in the $R$ band in the two high excitation regions (NGC\,2363 and 5253), 
due to the presence of strong \oiii\ lines. However, these estimates heavily 
depend on the
assumed shape of the passband, because of the location of the 
\oiii\ lines in the steeply falling part of the $B$ and $V$ band responses.
We find that even a 50\AA\ shift of the $V$ band response curve to longer
wavelengths changes the $B-V$ and $V-R$ colors by as much as 0.3 mag in
NGC\,2363. This amount of shift of the effective wavelength is not unusual and
hence gaseous corrected cluster colors using broad bands are bound to have
large errors for high excitation regions. Use of narrower bands situated
at emission line free regions would have eliminated this source of
uncertainty. 

The estimated nebular luminosities are subtracted from extinction
corrected broad band luminosities to obtain pure cluster quantities. These
quantities are reddened back to obtain the observationally expected values
without any contamination from the surrounding nebula.
We denote these quantities by the subscript $\ast$, 
while the observed quantities are denoted by the 
subscript `obs' (\eg\ $(B-V)_{\rm{obs}}, (B-V)_\ast$ \etc). The resultant 
quantities are corrected for the galactic and total extinction using $A_v$(gal) 
and $A_v$ and are denoted by suffixes 1 and 2  respectively. 
The Galactic extinction 
curve as tabulated by Seaton (1979) is used in these computations.
The process is summarized below. 
\begin{eqnarray}
 (B-V)_{\rm{obs}} &\rm{--- gas\ subt} \longrightarrow (B-V)_{\ast},  \\
(B-V)_{\ast} &\rm{--- }A_v(\rm {gal}) \longrightarrow (B-V)_1  \  ,  \\  
(B-V)_{\ast} &\rm{--- }A_v(\rm {Bal}) \longrightarrow (B-V)_2  \  ,
\end{eqnarray}
Table~2 contains the results for individual \hii regions. 
Column 1 contains identification numbers for GEHRs following Paper~I.
Oxygen abundances in column 2 are derived from \ohb\ ratio 
using the semi-empirical calibration of McCall \etal\ (1985). Note that
the sample regions have metallicities in the range 2--4\zsun, with the 
exception of the two high excitation regions mentioned above. McCall \etal\ 
also provide relation between the oxygen abundance and $T_e$,
which is used to derive $T_e$ given in column 3. Column 4 lists 
the visual extinction based on the Balmer decrement. Oxygen abundance
and visual extinction are based on the sources of spectroscopic data listed
in Tab.~1. While deriving the visual extinction, the 
underlying absorption in Balmer lines, which at the most 
contributes about 2\,\AA\ in \hb\ equivalent width (McCall \etal\ 1985),
has been accounted for. The $V_\ast$ magnitude, and $(B-V)_\ast$ and 
$(V-R)_\ast$ colors in columns 5, 6 and 7 respectively, are cluster 
quantities after subtracting the gaseous contribution. Column 8 contains the 
$\log$(\philb)$_\ast=\log(\phi)-\log{(L_B)}_\ast$ values. Note that 
$A_v$(Bal) has been used in obtaining the extinction corrected $\phi$ from
observations. Lyman continuum luminosity is given in column 13. Remaining 
columns in the table will be explained in the  next section.

\section{Comparison with Models}             

We plot in Fig.~1 the values of $\log$(\philb)$_1$ against $(B-V)_1$, 
which are corrected only for galactic extinction. Fig.~2 shows a similar
plot, but with $(V-R)_1$ replacing $(B-V)_1$. 
\hii regions from different galaxies are shown by different symbols, as 
indicated at the right hand side of the diagram. The small and large 
aperture measurements are distinguished in the case of NGC\,2997. Rms 
errors on the plotted quantities are indicated by the cross at the bottom-left 
corner of the diagram. For the two high excitation regions, NGC\,2363 and
5253, there is an additional source of error, arising out of the subtraction
of nebular contribution, as discussed in the previous section.
The arrow on the right side of the figure is the reddening vector, with its 
length corresponding to 1 magnitude of visual extinction. 

The observed quantities are compared with the evolutionary population 
synthesis model, which is briefly described below. The computational details and
results for different star formation scenarios can be found in Mayya (1995: 
Paper~II henceforth). The evolutionary population synthesis
model is based on the input stellar evolutionary data of Schaller {\it et al.}
(1992), and stellar atmosphere data of Kurucz (1992). The cluster is defined  
by an initial mass function, age, metallicity and the total mass in stars.   
The evolution of the cluster is performed at every 0.1~Myr interval, which is
adequate to sample even the fast evolutionary phases of massive stars. Optical 
and near infrared colors are computed and tabulated in Paper~II for three
different scenarios of star formation at solar metallicity. 
As noted earlier, GEHRs in the
present sample have metallicities between 2--4\zsun, with the exception of
NGC\,2363 and 5253 ($1\over 5$ and $1\over 3$\zsun\ respectively).
Recently, Cervi\~no \& Mas-Hesse (1994) and Leitherer \& Heckman~(1995), 
have shown the metallicity differences to affect the cluster evolution
significantly. 
Hence we extended the model computations of Paper~II to
Z$=0.1$\zsun\ and Z$=2$\zsun\ based on the stellar evolutionary tracks of 
Schaller \etal\ (1992) and Schaerer \etal\ (1993) respectively. Metal-poor
models are particularly important for the study the star formation in 
blue compact galaxies. Note that the metallicity of the most metal-poor galaxy,
\IZw\ is around a factor of 4 smaller than our metal-poor model.
The evolutionary tracks for an Instantaneous Burst (IB) of star formation 
at the three metallicities are shown
in Figs.~1 and 2. The long-dashed curve represents the solar metallicity model,
while the short-dashed and the solid curves represent 0.1\zsun\ and 2\zsun\ 
models respectively.  The clusters are evolved
upto 6.5~Myr, with dots placed on the curves at every 
0.5~Myr. The evolutionary phase corresponding to 4~Myr is indicated by the
letter 4 on each of the curves. The IMF chosen is close to Salpeter's: 
$m_l = 1$~\msun, $m_u = 60$~\msun\  and $\alpha = 2.5$. 

Two phases can be distinguished during the early (age$<$10~Myr) evolution of 
an IB. The first phase is controlled by the massive main sequence stars while 
the second by the supergiants, in particular the red supergiants. The 
transition from the first phase to second phase takes place around
4--5~Myr, with metallicity playing a major role during both the phases. 
The main sequence 
life time of the stars decreases with the increasing metallicity resulting in 
a faster drop of \philb\ in metal-rich regions during the first phase. Higher
metallicity favours the formation of red supergiants (RSGs) and Wolf-Rayet 
stars in the cluster and hence a given burst of star formation produces more 
of these stars in a metal-rich environment. Additionally the metal-rich RSGs 
are cooler than the metal-poor RSGs, resulting in redder colors during the red 
supergiant phase of metal-rich clusters. The increased optical luminosity
due to RSGs in metal-rich regions also leads to a decrease of \philb.
The metal-poor locus turns back at 5~Myr (\philb\ increases and colors
become bluer), which can be understood in terms of the absence of long
lived cooler supergiants in the cluster. 
Thus, the metal-rich clusters have lower values of \philb\ throughout their
evolution, and redder colors after around 4~Myr of evolution in comparison to 
metal-poor regions. A majority of the regions in our GEHR sample are metal 
rich and hence we use Z$=$2\zsun\ models for comparison with our data. 

%
For $m_u$ other than 60 (stellar masses are expressed in solar units
henceforth) the evolutionary track is similar to the one shown, except that 
the starting (age = 0) \philb\ value is different. For $m_u> 60$, it starts at 
a higher value than shown and the converse is true for lower masses. At 
$\alpha =2.5,$ the $B-V$ color reddens by 0.04 mag as the upper cut-off is 
decreased from 120 to 30\msun, whereas $\log$(\philb) decreases by 0.70~dex for
the same range of masses. As the more massive stars die, the evolution follows
that of a cluster with a lower $m_u$. 

It can be noticed that the evolutionary track forms an envelope around the 
observed points. All the observed points can be brought into the locus
expected from the Z$=2$\zsun\ model, by shifting the points along the 
reddening vector by amounts $A_v=$\,0--1.4~mag. The giant \hii complex
\ngc 2363 has several properties which distinguishes it from the
rest of the sample, and hence we treat this region separately in the
discussions to follow. The distinguishing properties are its low metallicity,
negligible extinction (Kennicutt \etal\ 1980), and a
large correction for the nebular contribution in $B,V$ and $R$ bands.

In Figs~3 and 4, the observed quantities are plotted after dereddening by 
$A_v$, as derived from Balmer decrement (quantities with suffix 2). Other 
details in these figures are similar to Figs~1 and 2, except that only
Z$=2$\zsun\ evolutionary track is shown and the track is truncated
at 6~Myr instead of 6.5~Myr. It is striking to 
note that the reddening correction has moved the points away from the models 
in both the axes. Hence the corrected $B-V$ and $V-R$ colors are too blue 
and (\philb) too low compared to the values expected for clusters younger than 
around 4~Myr. One can infer from Figs~1 and 2, that there are many regions for 
which very little extinction correction is required, whereas the extinction 
values computed using the Balmer decrement have a median value close to 1.5~mag,
with only 4 regions having values less than 0.75 mag. These two figures 
indicate that by dereddening the cluster related quantities by extinction 
derived from the Balmer decrement, we are over-correcting for extinction.
In order to investigate this in more detail, we plot in Fig.~5, the dereddened 
$(B-V)_2$ colors against the pure extra-galactic visual extinction 
$A_v-A_v$(gal). A trend of blue $(B-V)_2$ colors having higher extinction is 
clearly seen in the figure. A similar trend is also seen in $(V-R)_2$ color, but
to a lesser degree (not plotted). In fact both $(B-V)_2$ and $(V-R)_2$ colors 
are bluer than the model for regions having $A_v > 1.5$ mag. The figure gives 
an important clue on the nature of interstellar extinction towards stellar 
continuum and ionized gas emission in GEHRs. The inescapable conclusion from 
Figs~3, 4 and 5 is that the cluster continuum on an average experiences lesser 
extinction than the emission from the ionized gas. 
Such a situation is possible if most of the extragalactic extinction is due 
to the dust internal to GEHRs and the radiation from the ionized gas is
selectively absorbed by the dust. The ionized gas in GEHRs is known to
be confined to clumps, with a volume filling factor as small as 0.01
(Shields 1990). The dust is also likely to be present in the same volume.
Radiation from the stars, which contribute to the optical and UV 
continuum, on the other hand might occupy less dustier regions, 
and hence experience lesser amount of extinction. The segregation might be
arising as a result of relative motion of the stars with respect to the
gaseous material from which they have formed (Leisawitz \& Hauser 1988).
Based on this simple clumpy model, we formulate 
a different prescription to estimate the extinction of the stellar continuum 
which should be more accurate at least statistically. 

\subsection {Estimation of Extinction towards Cluster Continuum }  

The observed range of colors is too large to be explained in terms of
IMF variations. For an assumed slope of the IMF, the $B-V$ color 
reddens by $\leq 0.04$ mag as $m_u$ decreases from 120 to 30\msun. Color 
differences are around 0.20 mag even for two extreme IMFs, one rich in massive 
stars ($m_u=120$, $\alpha=1.0$), and the other rich in low mass stars 
($m_u=30$, $\alpha=3.5$). Thus, as a first approximation, we can neglect the 
dependence of colors on the IMF for clusters younger than 3.5~Myr. We assume 
$B-V=-0.25$ and $V-R=-0.10$ as typical values for the young clusters in our 
sample, corresponding to $\alpha = 2.5.$ and $Z=2$\zsun. The remaining part of 
the discussion pertains to the IMF parameters $m_l=1, m_u=60, \alpha=2.5$, 
unless otherwise mentioned. The regions having high \philb\ value in addition 
to blue $B-V$ and $V-R$ colors must be the youngest regions in the sample.
We hence make independent estimate of visual extinction ($A_{vc}$) towards 
stellar continuum on the basis that dereddened $B-V$ and $V-R$ colors cannot be
bluer than $-0.25$ and $-0.10$ respectively. The method is summarized
by the following equations, which are based on the formulation of Calzetti
\etal\ (1994). 

Consider an embedded cluster with intrinsic intensity $I_\lambda^0$ at a given
wavelength. The intensity $I_\lambda$ leaving the parent galaxy
is then given by,
\begin{equation}
   I_\lambda = I_\lambda^0  \exp (-\tau_\lambda)  \ ,
\end{equation}
where $\tau_\lambda$ is the line of sight optical depth suffered by the 
radiation at wavelength $\lambda$ in the host galaxy.
If $\tau_\beta$ and $\tau_\alpha$ are the optical depths experienced by the
Balmer line photons \hb\ and \ha\ respectively, then the differential optical 
depth between \hb\ and \ha\ is given by,\\
\begin{equation}
\tauBl = \tau_\beta - \tau_\alpha 
= \ln \left({{(H\alpha / H\beta)_i}\over{2.86}}\right) \ , 
\end{equation}
where (\ha/\hb)$_i$ is the ratio of intensities of \ha\ and \hb\ lines after
correcting for galactic extinction. Visual extinction $A_v$ tabulated in
Tab.~2 is related to $\tauBl$\ by,\\
\begin{equation}
A_v(\rm{Bal}) - A_v(\rm{gal}) = {{1.086\times 3.1}\over {X_\beta - X_\alpha}}
\tauBl \ ,
\end{equation}
or
\begin{equation}
\tauBl = 0.354 (A_v(\rm{Bal}) - A_v(\rm{gal})) \ ,
\end{equation}
where $X_\beta$ and $X_\alpha$ are tabulated by Seaton (1979) for the galactic
extinction curve. $\tauBl$ as defined above is the difference in extinction 
between \hb\ and \ha\ emission lines. Analogously we can define the extinction
suffered by the continuum photons between \hb\ and \ha\ wavelengths. We denote
this quantity as $\tauBc$ defined as, \\
\begin{equation}
\tauBc = \tau_\beta^c - \tau_\alpha^c 
= {{X_\beta - X_\alpha}\over{1.086}} E(B-V)_i   \ ,
\end{equation}
where \\
\begin{equation}
E(B-V)_i = (B-V)_{\rm{obs}} - (B-V)_{\rm{mod}} - A_v(\rm{gal})/3.1.
\end{equation}
We define,\\
\begin{equation}
g = {\tauBc\over\tauBl}
  = {{A_{\rm{vc}} - A_v(\rm{gal})}\over{A_v - A_v(\rm{gal})}}\ .
\end{equation}

Equations similar to the above are also formulated for the determination of
$g$ from the observed $V-R$ colors. There are 11 regions in the sample for
which both $(B-V)_2$ and $(V-R)_2$ are bluer than the bluest model colors. 
We derive a mean value of $g=0.70\pm0.20$, from both these colors. 
Physically $g$ is
controlled by the spatial distribution of gas, dust and stars with a value of
unity if all the three components are well mixed inside the volume of GEHR
or the entire dust is situated far away from the emitting region. However
there are several issues which come in the way of an accurate determination
of $g$, some of which are intrinsic to the region of interest and hence
unavoidable, while others are due to observational errors on colors.
In estimating $g$, we assumed all regions whose dereddened colors become bluer 
than the bluest model colors to be young. However a moderately evolved 
region with a lower value of $g$ than estimated is indistinguishable from a 
young region with $g$ value closer to unity. Thus the actual value of $g$
could be smaller than the estimated mean. In estimating $g$, we also assumed
that the entire extra-galactic extinction comes from within the star-forming
complex. In reality, there may be some contribution from the ISM of the 
parent galaxy. Correction to this has the effect of decreasing the
value of $g$. With all these considerations, we assume $g=0.50$ to be typical 
for the whole sample of GEHRs in the subsequent analysis. 
This value is supported by a recent analysis involving UV continuum and 
Balmer lines of central regions of galaxies (Calzetti \etal\ 1994).
Using this average value, $A_{vc}$ is 
recomputed for all the regions using eq.~12. The 
resultant values are tabulated in column 9 of Tab.~2. The cluster quantities 
are dereddened using these values of $A_{vc}$ and the resulting quantities are 
identified by suffix 3. Nebular quantities and hence $\phi$ are corrected for
extinction using $A_v$. 

\begin{equation}
(B-V)_{\ast} \rm{--- }A_{vc} \longrightarrow (B-V)_3
\end{equation}

The resultant values of (\philb)$_3$, $(B-V)_3$ and $(V-R)_3$ are 
tabulated in Tab.~2. $\log$(\philb)$_3$ is plotted against $(B-V)_3$ and
$(V-R)_3$ in Figs~6 and 7 respectively. As in the earlier figures the
IB model for the chosen IMF is shown with solid curve, with dots placed
every 0.5~Myr. The reddest point on the curve corresponds to 6~Myr of
evolution. With the new prescription for extinction correction, not only the 
colors, but also the $\log$(\philb) values are brought closer to the curve 
represented by the IB model. While a set of regions follow this track in both 
the diagrams within the observational errors, there are some regions which lie 
below and some others which lie to the right (redward) of the solid line.
These regions require alternative models, which is the main theme of
the discussion in the remaining part of this section.

\subsection{Destruction and Escape of Ionizing Photons}   

There are statistically significant number of regions which have blue
colors, indicative of young age, but have \philb\ values lower than that
expected for a cluster younger than 3~Myr. These regions seem to agree
well with the model, if the $\log$(\philb) is shifted by about $-0.2$~dex,
as indicated by the dashed curve in Figs~6 and 7. The possible sources, which 
can give rise to this shift are investigated below.
We express the discrepancy in $\log$(\philb) between observations and the 
model evolutionary track by a parameter $f$ defined as, 
\begin{equation}
\log({\phi\over{L_B}})_{\rm{obs}} = 
\log({\phi\over{L_B}})_{\rm{mod}}(t) + \log f,
\end{equation}
where (\philb)$_{\rm{obs}} = $\,(\philb)$_3$ and 
$\log$(\philb)$_{\rm{mod}}(t)$ is the IB model value at age $t$. 
For the IB model,
changes in \philb\ are mainly controlled by the changes in $\phi$ and hence,
physically, $f$ is the ratio of the detectable number of Lyman continuum 
photons to that expected from the model and has a maximum value of 1
when every Lyman continuum photon emitted by the cluster stars at the
estimated age can be traced by the Balmer emission lines. We estimate ages
between 2.5--4~Myr for the regions which require $-0.2$~dex shift. This value 
of shift implies a value of $f=0.65$ or 35\% of the ionizing photons could not 
be traced by the \ha\ emission. 

The following processes may be responsible for the missing ionizing photons:
(i) the destruction of ionizing photons by dust within the ionized complexes 
($\eta_{\rm{dust}}$), and 
(ii) the escape of ionizing photons from the nebula or an under-estimation of 
\ha\ flux, which traces the ionizing photon rate ($\eta_{\rm{esc}}$).
The under-estimation may be caused by the total
absorption of \ha\ photons by dust clumps or because a significant fraction
of the total \ha\ luminosity is in low surface brightness halo component,
which lies below the detection limit in most cases. 
$\eta_{\rm{dust}}$ and $\eta_{\rm{esc}}$ are related to $f$ by the eq.,
\begin{equation}
 f = (1-\eta_{\rm{dust}})(1-\eta_{\rm{esc}}).
\end{equation}
Good estimates of $\eta_{\rm{dust}}$, $\eta_{\rm{esc}}$ are lacking both
observationally and theoretically. Smith \etal\ (1978) and Mezger (1978) have 
estimated $\eta_{\rm{dust}}$ to be between 0.7 and 0.2 for giant
galactic \hii regions. Belfort \etal\ (1987) have assumed
a useful mean value of $\eta_{\rm{dust}}=0.30$, a value also used by 
many others (see \eg\ Mas-Hesse \& Kunth 1991). The observed value of 
$f=0.65$, then corresponds to negligible $\eta_{\rm{esc}}$.
The value of $f$, however, depends on the assumed IMF parameters 
with the dependence weakening beyond 3~Myr.

From these analyses, it may be concluded that the absorption of ionizing photons
by dust is the main source of the discrepancy in \philb\ (or Balmer line 
equivalent widths) observed in some regions, for the assumed IMF parameters. 
This is also supported by our conclusion based on cluster colors, that dust 
resides inside GEHRs. 
On the other hand a non-zero $\eta_{\rm{esc}}$ cannot be completely ruled out 
because of the uncertainties in $m_u$, age and absorption fraction of ionizing 
photons by dust. 
The detection of extra-\hii region ionized gas in several irregular galaxies by
Hunter \& Gallagher (1992) and Hunter \etal\ (1993) has raised questions
regarding the source of ionization of these diffuse regions. The Lyman
continuum photons escaping the \hii complexes may explain the observed 
ionization.

\subsection{Effect of Multiple Star Bursts }   

The evolution of an instantaneous starburst with around 35\% loss of the Lyman 
continuum photons, along with reduced effective extinction towards radiation 
from stars compared to that from the ionized gas, is found to explain the 
blueward envelope (within the limits of observational scatter) of regions in the
$\log$(\philb) vs $(B-V)$ and $(V-R)$ diagram. However, there is considerable
spread of colors at any given \philb\ which cannot be explained by
an instantaneous burst model. 

The departure of the colors from the IB model is in the redder direction,
which can be understood in terms of a population rich 
in red supergiants, co-existing with the younger population. 
The two populations of different ages are seen spatially resolved
in 30 Doradus (McGregor \& Hyland 1981). The younger population 
provides the ionizing radiation whereas the red supergiant population 
contributes bulk of the optical continua. Stars in the mass range
40--15\msun\ become red supergiants at ages 4--14~Myr at $Z=2$\zsun\  
(Schaerer \etal\ 1993). Simultaneous existence of ionizing stars as well as 
red supergiants implies extended duration of star formation either proceeding
continuously or in the form of bursts of short durations. The range of
computed quantities for continuous star formation is small compared to
the range of observed quantities (Paper~II), and hence the continuous star
formation cannot explain present observations. On the other hand, 2-burst 
models of Paper~II reproduce much wider range of observable values. We discuss 
these models in order to explain our observations. 

The composite model of two starbursts of equal strength with the 
younger one at 0, 3 and 4.5~Myr is shown in Figs~6 and 7 as three sequences 
of short-dashed lines. Both the bursts assume IMF slope of 2.5 and $m_u=60$ and
$Z=2$\zsun. In each of these sequences, the age of the older burst changes from
6--14~Myr, with beginnings of the sequences marked by the short vertical
line followed by successive solid dots spaced every 0.5~Myr. The color
for IB model becomes reddest at around 7~Myr corresponding to the RSG phase of 
stars around 25\msun\ mass, turning blue again at later stages of
evolution. This is the cause of the blue-ward turnover of the 2-burst
models. The general trend of many observed regions lying redward of the IB 
model is well reproduced by the 2-burst models, with younger burst ages
between 0 and 4.5~Myr and the older burst in the supergiant-rich phase
(age $\sim 10$~Myr). 

For the regions with two bursts of star formation, the observed values of
$\log$(\philb) can be reproduced by models without the need for any process 
that significantly destroys the ionizing photons, unless
the upper cut-off mass of IMF is higher than 60\msun.
In Figs~6 and 7, the tip of the vertical arrow indicates the position of IB 
model at $t=0$ if $m_u$ is increased to 120\msun. 

\subsection{Classification of Sample GEHRs }      

Based on the position of the GEHRs in $\log$(\philb)$_3$ vs $(B-V)_3$ and 
$(V-R)_3$ diagrams we classify the GEHRs into 5 groups. The classification is 
uncertain for those few regions with a poor correspondence between $B-V$ and 
$V-R$ colors. Nevertheless the classification 
is expected to throw more light on the evolutionary history of GEHRs in 
general. A short description of the five groups are given below. \\
1. Young regions ($t_{\rm{young}}<3$~Myr) with instantaneous burst of star 
formation.\\
2. Moderately evolved regions ($t_{\rm{young}}\sim$\,3--5~Myr) 
with instantaneous burst of star formation.\\
3. Young regions ($t<3$~Myr) with a co-existing population of red 
supergiants from a previous generation.\\
4. Moderately evolved regions ($t_{\rm{young}}\sim$3--5~Myr) with a 
co-existing population of red supergiants from a previous generation. \\
5. Evolved regions ($t>5$~Myr) rich in red supergiants containing stars
from one or two bursts. \\

\noindent
Table~3 lists the objects classified under each group. None of the regions
in our sample can be classified into group 5. The last row in the table gives 
the percentage of observed regions in each group. A colon in front
of an entry suggests that the classification is uncertain either because
of lack of correspondence between $B-V$ and $V-R$ colors or due to more
than one model occupying a given region in our diagnostic diagrams
(within the observational errors). The classification helps us to draw the
following conclusions.\\
1. Seventy percent of the sample regions contain stars from more than one
burst over the last 10~Myr. The most recent burst has occurred within the
last 3~Myr in 90\% of these regions.\\
2. The frequency of detection of young single burst regions is low,
implying that the first generation of stars might be occurring
deep inside the clouds, and hence are optically obscured when they are
young and luminous.\\
A detailed scenario of star formation, incorporating all the 
observed properties is proposed in section~5. 

\subsection{Masses of Embedded Clusters }      

For an assumed IMF, the mass of the star-forming complexes can be estimated
based on the observed Lyman continuum luminosity of the regions. Corrections 
for the decrease in ionizing luminosity due to evolution, destruction by dust 
and possible escape of photons have to be made in order to obtain the total 
ionizing luminosity emitted from the cluster. The total mass ($M_T$) is thus 
given by,
\begin{equation}
 \log(M_T) = \log(\phi)_{\rm{obs}} - \log f - \log(\phi)_{\rm{mod}}(t),
\end{equation}

\noindent
where $(\phi)_{\rm {mod}}(t)$ is the ionizing luminosity per unit mass of
the cluster at an age $t$, as derived from the model for an assumed IMF. 
The factor $f$ takes care of 
the loss of ionizing photons with an estimated value between 0.65 and 1.0. 
A mean $(\phi)_{\rm{mod}}(t)$ is assigned for each group (Tab.~3), which is 
used in deriving the mass of individual regions. We further assumed that only 
the younger burst contributes ionizing luminosity, when two populations 
co-exist. The masses derived with these assumptions and an IMF with 
$\alpha=2.5$ and $m_u=60$ are given in the last column of Table~2. Their 
histogram distribution is given in Fig.~8 with a mean value of
$$ \log(M_T) = 5.1 \pm 0.5 $$
for the entire sample ($n=29$). While estimating the masses we assumed $f=1$,
implying complete detection of ionizing photons. Depending on the exact value
of $f$ for a given region, the actual masses might be higher by a
factor of up to 2. The above estimates indicate that the mean mass of a cluster
is $\sim 10^5~$\msun. The masses range by a factor of ten on either side
of this mean value. The high mass end of galactic Giant Molecular Clouds 
(GMCs) reaches values up to $10^6$\msun\ (Solomon \etal\ 1987). 
From high resolution CO observations of molecular clouds in the arm 
of M\,51, Vogel \etal\ (1988) derive masses up to $3\times 10^7$\msun, 
referred as Giant Molecular Associations (GMAs). Elmegreen~(1994)
has recently addressed the formation mechanism of such clouds and
found that $10^7$\msun\ clouds can be formed due to gravitational
instabilities in a spiral arm before the gas flows out of the spiral arm.
Thus, depending on the mass of the parent molecular cloud, about 1--10\% 
of the available mass is converted into stars in each episode of star
formation. 

\subsection{Comments on Selected GEHRs }     

NGC\,2363 is left out from most of the above discussions because it is
different from the rest of the regions in the sample in terms of its low 
metallicity and extinction. This region was of special interest in many 
previous investigations (Peimbert \etal\  1986, Kennicutt \etal\ 1980). This 
is one of the regions with the smallest Balmer decrement values. The absence of 
2200\,\AA\ absorption feature in the IUE spectrum (Rosa \etal\ 1985) also 
suggests a low interstellar extinction towards this region. \philb\ for this
region is the highest in the sample. The estimated age is less than
2~Myr, assuming an IMF with slope of 2.5 and $m_u=60$. Broad band
colors for this region are highly contaminated by the strong emission lines
($>40$\% in $V$ and $R$ bands), and hence an investigation regarding the 
existence of the second population cannot be carried out. Drissen \etal\ (1993)
find the kinematic age of the expanding bubble in this region to be
2~Myr. They also found Wolf-Rayet (WR) stars suggesting ages around 
3~Myr. It is likely that $m_u$ was close to 120 in this region,
with the most massive stars in their WR phase. 
The inferred mass of the cluster is $4\times 10^5$\msun. 

The starburst nucleus of \ngc 5253 has also been well studied using
multi-wavelength observations (Walsh \& Roy 1989a; Gonzalez-Riestra \etal\ 
1987; Moorwood \& Glass 1982). The region has sub-solar metallicity.
Emission lines contribute around 15--25\% in the $BVR$ bands. The region 
agrees very well with the 2-burst model, with the second population just
formed. The inferred mass of the embedded stars is $10^5$\msun. 

Among the GEHRs in spiral arms, the most interesting complex is the one lying 
at the tip of the north-east arm of \ngc 2997(N2997NE-G1). The complex 
contains three knots (denoted by 1, 2, 32 in Paper~I) of similar brightness 
in the continuum light while the knot 32 is totally absent in the \ha\ image.
Knots 1 and 2 are treated as independent regions and are included in our plots. 
We infer that the knot 1 itself contains stars from more than one burst, with a 
very young second burst. 
The bigger aperture measurements containing all the three knots support the
results on individual knots. A similar behaviour is also seen in complexes 
N2997NE-G3 and N2997NE-G4. 

\subsection{Emission line and continuum peaks of GEHRs}   

In two of the program galaxies (\ngc 2997, 4303), we found a noticeable 
mismatch of continuum and line emitting peaks significantly larger than the
alignment errors. These results are particularly worth the attention,
as there was a good agreement of the positions of knots in the $BVR$ bands. 
As noted in Paper~I, this mismatch has necessitated us to choose
centers and radii of apertures by examining both the broad band and
\ha\ images. We attempted getting a rough estimate of the separations between 
the emission line and continuum knots in two of the galaxies mentioned above. 
The measured separation between the peaks of nebular and stellar continuum 
knots is $42\pm20$~pc for 5 regions in \ngc 2997 and $100\pm37$~pc for 13 
regions in \ngc 4303, respectively. It is obvious that the
lower limit to the separations is due to the lack of resolution ($\sim$ 18 and
34 pc at the distances of \ngc 2997 and 4303). The upper limit to the 
separation comes from the typical separation of star-forming complexes 
(chain of \hii regions) in spiral arms. The mean separation is $450\pm100$~pc 
and $1200\pm475$~pc for the two galaxies that we have considered. It is 
interesting to note that the separation between the neighbouring knots with 
striking brightness differences are around one tenth of the mean separation 
between distinct \hii complexes. Also, the separation between the nebular and 
continuum knots that we have obtained are similar to the sizes of the
largest Giant 
Molecular Clouds. This separation is also of the same order as the one between 
the old and the young populations in 30 Dor (Hyland \etal\ 1992).

In order to explain the photometric properties of \hii complexes, we required a
population older than 5~Myr superposed on a younger population for a majority of
the sample regions. We are tempted to believe that in some cases we have
actually resolved these two populations  as \ha\  and continuum knots.
If the older population has triggered the formation of younger population, 
we estimate the speed of propagation of the trigger as 40--100~pc in 
$\sim10$~Myr, or 4--10~km\,s$^{-1}$. This  value is reasonable for a 
trigger in terms of mass-loss driven stellar wind from the older population.

Direct confirmation of these results will require spectroscopy of continuum 
bright knots for detectable signatures of red supergiants. Images taken at 
high spatial resolutions, such as possible with the Hubble Space Telescope will
be invaluable in quantifying the separations with reduced uncertainties.

\subsection{Summary of Results }     

From the discussions above, we draw the following conclusions regarding the 
extinction and star formation in GEHRs in our sample. \\
(1) The dereddened colors of the embedded clusters are found to correlate 
with the visual extinction estimated from the Balmer decrement. Regions having 
the bluest colors have larger $A_v$ values, implying an over-estimation of 
reddening towards embedded cluster stars. The fraction $g$ 
of visual extinction experienced by the stars to that by the nebular gas
is estimated to be $0.7\pm0.2$, from a subset of the bluest regions in the
sample. The value of $g$ is likely to be even smaller if one considers the
various uncertainties in the determination of $g$. We use a value of
$g=0.5$, which is the value derived by Calzetti \etal\ (1995) for nuclear
star-forming regions.\\
(2) GEHRs with a single instantaneous burst of star formation are rare
in our sample. Loss of ionizing photons amounting to $\sim 35$\% is inferred
in some regions, which may be explained as caused by the direct absorption
 by dust,
if $m_u=60$. Other mechanisms such as the escape of ionizing photons through 
lower density regions, or total absorption of some of the Balmer line photons 
by intervening dust clouds, have to be invoked if $m_u>60$. 
Escaping Lyman continuum photons possibly ionize the low
density inter-stellar medium, thus explaining the origin of extra-\hii
region ionized gas detected by Hunter \& Gallagher~(1992).  \\
(3) There is a large scatter in the $B-V$ and $V-R$ colors for a given
\philb. Adding a population rich in red supergiants, supposedly from an
earlier burst, to a young population responsible for ionization, 
explains all the three observables. 70\% of the sample regions are found
to be in this phase of star formation. The estimated age of the older 
population is 10~Myr. \\
(4) We find the \ha\ emitting knots spatially separated from the continuum
knots in \ngc 2997 and 4303, from careful measurements of positions of 
knots on \ha\ and broad band images. If we attribute these knots to
young and old populations, there is a spatial separation in addition to the 
age difference between the two populations. A mild trigger induced
by mass-loss winds from massive stars of previous generation is a viable 
mechanism for producing successive generations of starbursts in GEHRs. \\
(5) The evaluation of IMF parameters namely $m_u$ and $\alpha$, which are
greatly needed in understanding star formation processes in different
environments, is hampered by the uncertainties in $g$, in quantifying 
the loss of ionizing photons and by the
existence of an older population. The present observations do not compel us
to choose an IMF slope far away from the Salpeter's value of 2.35. An upper
cut-off of 60\msun\ can explain a majority of the regions.\\
(6) The mass of gas converted to stars in each burst is $10^5$\msun, with an
order of magnitude spread in mass on either side. \\
(7) The sample contains many metal rich regions. It is desirable that
stellar evolutionary codes become available at metallicities higher 
than 2\zsun, considering the heavy dependence of evolution on 
metallicity. \\

\section{Discussion}   

In this section we discuss the main results obtained in our work,
namely the reduced extinction towards stellar continua and the co-existence of
two populations in a majority of the GEHRs, in the context of earlier work on 
GEHRs and related objects.
There have been studies of GEHRs and \hii galaxies that show indications of
reduced extinction towards stars, although interpreted differently by the
authors. We look at those results afresh in the following paragraphs. 
 
\subsection{\hb\ equivalent widths }     

The most widely used diagnostic parameter for star formation so far is the 
spectroscopically derived \hb\ equivalent width.
Viallefond (1987), from his compilations, finds that the \hb\  equivalent
widths and the effective temperatures of the cluster (defined by the ratio of
helium ionizing to hydrogen ionizing photons) cannot be simultaneously
explained. He finds the observed \hb\  equivalent widths to be lower by a
factor of 3--4 compared to models with Salpeter IMF without evolution.
This is essentially the same discrepancy that we have found in Figs~3 and 4,
where observed $\log$(\philb) values are much lower than the IB models
with moderate evolution. Aperture size effects and usage of improper
model atmosphere were thought by Viallefond to be partly responsible for the
discrepancy. The method followed in the present study takes 
care of both these effects, but still retains the discrepancy. The data
set of Terlevich \etal\ (1991) on \hii galaxies also shows the discrepancy as
discussed below.

A spectrophotometric catalogue of \hii galaxies containing 425 objects
is published by Terlevich \etal\ (1991). It is known since the work of
Sargent \& Searle (1970) that \hii galaxies are ionized by massive clusters of
coeval OB stars. These galaxies are presently experiencing a high level of star
formation activity closely resembling ``starburst'' galaxies on
the higher scale and GEHRs on the lower scale. Observable properties are 
dominated by young stellar component with little or no evidence for any 
underlying population (Campbell \& Terlevich 1984; Melnick \etal\ 1985). 
Among other quantities, Terlevich \etal\ publish \ha\  and 
\hb\  equivalent widths in their catalogue. Like giant extragalactic \hii
regions these galaxies are also thought to be simpler systems dominated by a
young star-forming complex. Thus, the evolutionary population synthesis
model discussed in Paper~II is valid for these systems as well. As a
class, \hii\ galaxies are low metallicity systems and hence we compare the
median \ha\  and \hb\  equivalent widths of this sample with our model at
$Z=0.1$\zsun\ metallicity. 
The computed median \ha\  and \hb\  equivalent widths 
for the sample regions are 172\,\AA\ and 33\,\AA\ respectively. These values 
compare well with models which produce $\log$ (\philb)$_2$ values 
close to our median value of $-1.75.$ For the IMF with $m_u = 60$ and $\alpha =
2.5$ at $Z=0.1$\zsun, the IB model corresponds to evolution beyond 5~Myr.
Based on the photometry of Huchra (1977) and spectroscopy of Skillman \& 
Kennicutt (1993), we derive $\log$(\philb)$_2=-1.74$, $(B-V)_2=-0.02$
and $(V-R)_2=0.13$ for the blue compact galaxy \IZw. These values suggest 
an instantaneous burst of star formation of age $\sim5$~Myr, which implies that
the most recent burst in \IZw\ has properties similar to the average 
properties of \hii galaxies. 
Statistically one expects star-forming regions to be detected at their
bright young phase ($\le 3$~Myr). In the entire spectrophotometric
catalogue there are only three galaxies which have \hb\ equivalent widths 
$>350$\AA, the expected value for a cluster younger than 3~Myr. 

With our new prescription for extinction correction, \ha\ and \hb\
equivalent widths depend on the extinction, contrary to normal
convention. Correction to this effect increases the observed equivalent widths,
bringing them closer to the values expected for young clusters. The escape
of Lyman continuum photons from the nebula also leads to a decrease in the
observed equivalent widths. The addition of evolved stars from previous
 generations
adds to the stellar continuum, thus resulting in the decrease of Balmer line
equivalent widths. Thus, the observed low Balmer line equivalent widths in 
GEHRs and \hii galaxies can be explained by a combination of the above 
processes. 


\subsection{UV Continuum Observations }  

Young and massive stars emit most of their energy at UV wavelengths. Thus, the 
UV continuum serves as a very good tracer of massive star formation. The 
reduced extinction suffered by the stellar continuum, as we found
from optical colors, should be affecting the observed spectra in the UV
wavelength regions also. UV part of the spectra are essentially
independent of IMF differences and hence are good indicators of extinction.
Such spectra have become available only recently through rocket and satellite 
observations. Results from these UV spectra of GEHRs and other objects are 
discussed in the following paragraphs.

Rosa \etal\ (1985) have published UV spectra of GEHRs based on the
{\it International Ultraviolet Explorer} (IUE) observations. It is found that 
often the 2200\,\AA\ absorption feature is weak or absent, although the optical
extinction combined with the galactic extinction law predicts much stronger 
bump. From the observed bumps in five objects, Rosa \etal\ (1985) infer
$E(B-V)$ to be not larger than 0.5 for GEHRs. It should be noted that the
$E(B-V)$ value obtained from this method measures the ``true'' stellar
extinction, while that derived from the Balmer decrement measures the extinction
towards the gaseous component. These observations again indicate lesser 
extinction towards stellar continuum compared to the nebular radiation.

M\,83 was one of the targets for rocket UV imaging by Bohlin \etal\ (1990).
The UV images of this galaxy were obtained at 1540 and 2360\,\AA\ and aperture
photometry was performed on 18 \hii regions to get monochromatic magnitudes
$m_{\rm{fuv}}$ and
$m_{\rm{nuv}}$ respectively. They have performed $V, B-V$ photometry based on
the images of Talbot \etal\ (1979). By fitting a straight line to
the reddened $m_{\rm{nuv}} - V $ and $m_{\rm{fuv}} - V $ colors, they
find that the reddening curve towards stars in this galaxy is the same as
that for our galaxy. The color excesses $E(B-V)$ are derived by comparing 
the observed $m_{\rm{nuv}} - V $ colors with a model with Salpeter 
slope and $m_u=120$, $m_l=1.8.$ The mean value of color excess so derived is
$E(B-V)=0.25$ with a small scatter (0.05 mag rms). Dufour \etal\ (1980) present 
optical spectrophotometry of six \hii regions in this galaxy apart from the 
nucleus. The derived $E(B-V)$ values from the Balmer decrement have a median 
value of 0.75, three times larger than the values estimated from the UV 
continuum. These observations are again consistent with our findings that the 
reddening towards stellar component is less than that for the nebular 
component. Dereddening the colors with the $E(B-V)$ values of Bohlin \etal\ 
fits with the models with evolution within 2~Myr. 

Blue compact galaxies have attracted considerable attention in the UV, because 
of their expected blue spectrum. Like the study of GEHRs in M\,83, these 
galaxies also show much lesser extinction towards the UV stellar continuum 
compared to the values expected from the Balmer decrement (Fanelli \etal\ 1988).

Calzetti \etal\ (1994) have recently analysed IUE-UV and the optical 
spectra of 39 starburst and Blue Compact Galaxies in order to study the 
average properties of dust extinction in extended regions of galaxies. 
To date, this is the most exhaustive study on the topic of extinction towards
the stellar continuum and nebular radiation in star-forming regions, and hence 
is the most relevant from the point of view of our study.
Calzetti \etal\ find a correlation between the slope of the UV spectrum and the
optical depth between the Balmer emission lines \ha\ and \hb, which cannot
be satisfactorily reproduced by any of the five geometrical distributions 
of dust they have considered, for both Milky Way and LMC extinction laws. They 
derive an
extinction law directly from the data in the UV and optical spectral range. 
The resulting curve is characterized by an overall slope which is more
gray compared to the Milky Way extinction law, and by the absence of
the 2200\AA\ dust feature. With the new extinction law, the difference
in optical depths between \ha\ and \hb\ stellar continua is only half of 
the difference between the optical depths for the corresponding emission
lines. Our results on GEHRs, based on colors, are consistent with these
results. 

From the above discussions it is evident that the star-forming complexes share 
the same extinction properties irrespective of spatial scale and strength of 
star formation or position of the star-forming site in a galaxy. As a rule the 
stellar continuum in optical and UV regions suffers lesser amount of 
extinction compared to the radiation from the surrounding nebula. It is likely 
that the same physical process controls the extinction in all star-forming 
regions.

\subsection{Thermal Radio Emission}  

The extinction towards a nebula can be derived by using the observed intensity
ratio of thermal radio emission to Balmer lines ($A_v$(radio)). Most often 
the extinction is measured using two or more of the Balmer lines ($A_v$(Bal)).
Several groups have compared the extinction derived from these two
methods (Israel \& Kennicutt 1980, Viallefond \etal\ 1982, 
Caplan \& Deharveng 1986). General conclusion from these studies
is that $A_v$(radio) is around 1 mag greater than $A_v$(Bal).
Practically it is often difficult to match the apertures for radio and
optical measurements, which might explain a part of the uncertainty in the
above conclusions. However, it is believed that the observed higher value of
$A_v$(radio) is real and may be related to the distribution of dust between the
source and the observer. Caplan \& Deharveng (1986) investigated various dust
configurations to explain radio and optical extinction in a sample
of \hii regions in the LMC. From these they conclude that a uniform extinction 
cannot explain the LMC data. Some of the extinction is due to clumped dust, 
well outside of the emission zone, and the rest is caused by dust located
closer to or mixed with the ionized gas, where scattering effects are
important. Measurement of extinction directly from resolved stars
indicates that there is interstellar dust, quite irregularly distributed. 
Higher extinction from radio measurements implies that some of the
Balmer photons are totally absorbed by the intervening dust. Under these 
conditions, the Lyman continuum luminosity as measured from Balmer lines 
underestimates the actual ionizing radiation. The fraction $f$ (or more 
precisely $\eta_{\rm{esc}}$ defined in section 3.2)  
has a contribution from these undetectable \ha\ photons. The absolute
value of $f$ critically depends on the chosen $m_u$ and hence the estimation
of the missing number of ionizing photons cannot be made accurately even
if one derives Lyman continuum luminosities from thermal radio
continuum.

\subsection{The 30 Doradus Complex in LMC}   

The Tarantula or 30 Dor nebula in the LMC is the closest example of a GEHR. Its
relatively nearby location has helped its observation in a great detail. The
near infrared observations revealed the existence of two populations of stars
(red and blue) in this complex (McGregor \& Hyland 1981). More recently Hyland
\etal\ (1992) have surveyed the region to deeper limits ($K = 13$~mag) and 
the earlier results are confirmed. The red population is explained to be 
corresponding to the red supergiant phase of stars with mass $> 20$\msun\
formed about 20~Myr ago. The older population is spatially separated from the 
younger one by distances of the order of tens of pc. Hyland \etal\ suggest 
that the younger cluster has formed due to the collision of the parent cloud 
with the mass loss wind of the older cluster stars. The present work has 
brought out the evidence for such supergiants in a majority of extragalactic 
\hii regions. 

\section{Proposed Scheme of Star formation in GEHRs}  

In this section, we propose a scheme of star formation, consistent with
all the observed properties of GEHRs. The main results which the model tries to
explain are the following.
 
\subsection{Observational facts }  

\noindent
1. The brightest GEHRs in galaxies, often preferred for detailed spectroscopic
study, are very rarely found to be genuinely young with only one generation of 
stars.\\
2. Among the fainter GEHRs in galaxies, there are a few with very high \ha\
emission equivalent widths (see Paper~I). \\
3. \ha\ bright regions in galaxies often contain red supergiants, possibly
from an earlier burst ($\sim$10~Myr ago) of star formation.\\
4. The optical and ultraviolet continua from stars experience lesser amount of
extinction compared to radiation from the ionized gas.\\
5. Some of the neighbouring emitting knots have  
marked variation in the ratio of \ha\ to optical continuum. \\
6. On an average GEHRs contain about $10^5$\msun\ of stellar mass in each
burst. \\

\subsection{Proposed model }  

A scenario of propagating star formation in galactic molecular clouds was
discussed by Lada (1985). We extend this model to a larger scale to
explain the star formation process in GEHRs. To start with, we summarize
the known physical properties of GEHRs and the parent molecular clouds.
GEHRs show a core-halo morphology in emission lines, with densities
decreasing from around 100~cm$^{-3}$ in the core to 1~cm$^{-3}$ in the halo.
Kennicutt (1984) points out that the density in the halo almost equals the
 density
of the general ISM. The average sizes of core and halo are around 75~pc and 
250~pc, respectively (Sandage \& Tammann 1974). In comparison, Galactic GMCs 
from which stars form have average dimensions of 60~pc (Solomon \etal\ 1987),
which is of the order of the dimension of cores of GEHRs. Interestingly,
 the Giant 
Molecular Associations found in the spiral arms of M\,51 (Rand 1993) have also 
dimensions matching that of halos, though their average densities are higher 
than the observed densities of GEHR halos. It is likely that GMAs contain a
few GMC-like condensations, surrounded by an extended low density
inter-clump medium, which get transformed into core and halo respectively
after massive stars are formed. An examination of individual regions indicate 
higher extinction towards higher \ha\ surface brightness regions (for example
see NGC\,5461 in McCall \etal\ 1985). Thus,the  core in an \hii complex 
contains more dust than the surrounding low surface brightness halo. Emission 
line equivalent widths of Balmer lines seem to be higher near the core
compared to the value obtained by integrating over the entire complex,
implying that the stars are spread over a larger spatial extent compared to
the line emitting cores of GEHRs (Roy \etal\ 1989). Stars, when born, are 
confined to the highest density regions, thus coinciding with the position of 
the \ha\ core. However, the relative velocity of stars with respect to the 
cloud make them diffuse out of the core into the lower density regions in a 
few million years (Leisawitz \& Hauser 1988).

With the above description of the physical properties of GEHRs and
molecular clouds, the overall process of star formation can be divided into 
four phases, each with a distinct observable signature. These four phases are
schematically depicted in Fig.9. The outer boundary denotes the parent
molecular cloud, whose dimension is around 300~pc. The GEHR core is shown by 
the circle, with the bigger concentric circle denoting the extent of the
ionized halo. Stars are denoted by asterix. The irregular pattern
indicates the density inhomogeneities in the parent cloud. The approximate
time period in each phase is indicated. \\

\noindent {\bf Phase 1: }
We assume the GEHRs to form in Giant Molecular Associations, which contain
several regions of higher than average densities. Some of these condensations
grow into GMC-like structures, the densest of which collapses to form stars.
This results in an ionization of the high density gas in the immediate 
vicinity, forming the core of the GEHR. Dust associated with the
molecular cloud offers considerable optical extinction, allowing only a few of
the young regions to be detected optically as faint \ha\ emitting knots with 
weak or no continuum. Regions with high \ha\ equivalent widths are in this 
phase of evolution. The large extinction when the cluster is young, makes its
optical detection very difficult. 
$\tauBc$ is expected to be equal to $\tauBl$ in the very early phase, with
a tendency for $\tauBc$ to decrease as some of the optically luminous
stars diffuse to low density regions. The ionization front is advancing
into low density regions. This phase might last for 3~Myr at the most.

\noindent {\bf Phase 2: }
There might be a net reduction in the extinction towards a GEHR in a few
million years, as the energetic events associated with star formation 
destroy the dust particles in the core. The reduced extinction makes these
regions optically visible for the first time. Embedded
cluster stars might experience additional reduction in extinction as more
and more stars diffuse out of the high density cores. By this time the
ionization front has also advanced past the high density regions of the
molecular cloud, thus creating an extended low density ionized halo, which
can be detected only through sensitive observations. This might lead to
an underestimation of the ionizing photons. However, emission
from the core still dominates the total Balmer emission from the region. 
Under this configuration, photons from ionized gas encounter more dust clumps, 
compared to stellar photons. Thus, the stellar continuum experiences 
lesser extinction compared to radiation from the ionized gas ($\tauBc<\tauBl$). 
This phase might continue up to around 5~Myr. The second sketch in Fig.~9 
depicts this stage.

\noindent {\bf Phase 3: }
The onset of this phase is characterized by the appearance of luminous
red and/or blue supergiant stars in the cluster, as the massive stars
evolve off the main sequence in around 5~Myr. The 
cluster appears brightest in the stellar continua from ultraviolet to 
near-infrared during this phase. This luminous period continues until all 
the massive stars go through the red supergiant phase ($\sim$12 Myr). In this
phase the region is very faint in \ha. 

Massive stars lose a large amount of mass and energy during their life time,
both on and off the main sequence. This results in an increased pressure 
on the gas left behind after the initial burst of star formation, thus
creating another density condensation in the same GMA. Evidence is
accumulating for the presence of left over molecular gas in the vicinity 
of GEHRs (\eg\ NGC\,604 in M\,33: Israel \etal\ 1990; Wilson \& Scoville 1992).
The shocked layer starts collapsing, eventually forming a new generation
of stars. The delay between the onset of collapse and generation of new
stars is given by (Elmegreen 1992),
\begin{equation}
t_{\rm{delay}} = 2.4 \left( {M\over {10^5}}\right)^{0.25}~~\rm{Myr.}
\end{equation}
For a cloud of mass $10^6$\msun participating in the collapse this
corresponds to $t_{\rm{delay}} = 4.3$~Myr. The arrow in the third panel
indicates that the cloud is collapsing. 

\noindent {\bf Phase 4: }
The end of phase~3 and the beginning of phase~4 are characterized by the birth 
of the second generation of massive stars  the same parent molecular 
cloud. Typically this happens at around 8~Myr, with the exact epoch depending 
on the mass of the parent cloud and the onset of collapse. During this phase, 
the region becomes brighter again in \ha. By this time the stars from the first 
generation have evolved, and the cluster itself has moved into regions of 
lesser extinction. In this phase, the bulk of observed stellar continuum 
originates from the older generation and hence suffers lesser extinction 
compared to the ionized gas situated in dense regions, thus continuing the 
trend shown in Phase~2. GEHRs are expected to be in this phase at about 10~Myr
since the beginning of the first generation of stars.
For typical propagation speeds ($\sim 10$~km\,s$^{-1}$), the separation
between the two populations amounts to $\sim 100$~pc. Thus the two generations 
of stars can be distinguished,under good spatial resolutions,
as neighbouring \ha\ and continuum (stellar) peaks. \\

In this study, we restricted the evolution of GEHRs up to the second
generation. In reality, star formation might continue until the parent
clouds of GEHRs are fully disrupted by the energetic events such as
multiple supernova explosions. It is likely that bigger systems like \hii 
galaxies and central regions of galaxies experience several bursts of star 
formation before the consumption of gas is complete. However, the process of
formation of multiple bursts might be different in these bigger systems. 
Recently, Korchagin \etal\ (1995) have constructed 
propagating star formation models for the nuclear hot-spots in galaxies 
and found the star formation lasting for a few tens of million years.

Before building on the above tentative model, it is important to establish the
different phases by elaborate observations. This would involve a detailed study
of selected high Balmer equivalent width regions to establish that they are
young and also that their stellar continuum experiences similar extinction
as the radiation from the ionized gas. Regions in the present sample with
 definite
indication of older generation should be observed for the signatures of red
supergiants. GEHRs should be imaged in the optical emission and continuum bands
with high spatial resolutions, in order to directly resolve the 
spatially separated generations of
stars, one bright in the emission line and the other bright in the stellar
continuum. As all these observations are well within the observing
capabilities of most medium-sized optical-infrared telescopes, we hope that
the above model will be put to test soon. 

\section{Conclusions}  

We find an evidence for reduced extinction towards stellar continuum compared to
the extinction derived from the Balmer decrement. Genuinely young GEHRs 
containing stars from only one generation seem to be very rare in the present 
sample. Instead, the young and moderately evolved regions contain
an older population rich in red supergiants. All these observational facts are 
put together resulting in a simple qualitative model, which can be tested by a 
detailed study of selected regions. In the model, GEHRs are obscured by the 
dust associated with the high density gas when they are young, and are 
detected only when the stars have moved out of the highest density regions. 
This leads to a lower effective optical depth towards the stellar continua 
compared to the emission lines from the ionized gas, whenever GEHRs are 
detected. The mass-loss wind from the massive stars
triggers a second burst of star formation in the parent cloud, providing a
fresh input of ionizing radiation. The presence of stars from two bursts and
reduced net extinction makes these regions easily noticeable on galaxy
images, and hence are preferred regions in most of the studies. 
Star formation in GEHRs stops after two or more bursts, probably because of 
disruption of the parent gas cloud. 

\acknowledgements

The suggestions made by the referee Marshall L. McCall has helped enormously in
improving the paper, especially in revising the proposed model for 
star-formation.

\newpage

\figcaption{ 
%
Observed GEHRs in $\log$(\philb)$_1$ vs $(B-V)_1$ plane. The nebular
contribution is subtracted from broad bands and extinction corrections are
done only for galactic dust. Typical rms errors are shown by the cross at the 
bottom left corner. The arrow indicates extinction correction corresponding to
$A_v = 1$~mag. The evolution of clusters at $Z=0.1, 1$ and 2\zsun\ are shown by 
the short-dashed, long-dashed and solid curves respectively. An IMF with
parameters $m_u=60$ and $\alpha=2.5$ is used. The dots on the line are 
spaced $0.5$ Myr apart with the dot corresponding to 4~Myr marked by the 
letter 4. }
%
\figcaption{
Observed regions in $\log$(\philb)$_1$ vs $(V-R)_1$ plane. See Fig.~1 for 
other details.}
%
\figcaption{
Observed regions in $\log$(\philb)$_2$ vs $(B-V)_2$ plane. Extinction 
corrections are done based on the Balmer decrement. Other details are as in 
Fig.~1, except that only $Z=2$\zsun\ model is shown and the evolution is shown 
only upto 6~Myr.}
%
\figcaption{
Observed regions in $\log$(\philb)$_2$ vs $(V-R)_2$ plane. See Fig.~3 for 
other details. }
%
\figcaption{
The $B-V$ colors after correction for extinction derived from the Balmer
decrement are plotted against $A_v$ corrected for galactic extinction.
Note the trend of bluer regions having larger $A_v$.}
%
\figcaption{
Observed regions in $\log$(\philb)$_3$ vs $(B-V)_3$ plane. Extinction 
corrections are done using $A_{vc}$ with $g=0.50$. Solid line represents
the $Z=2$\zsun\ model evolved upto 6~Myr. The same model
is shifted by 0.20~dex (see text for details) and shown by the dashed line.
A model with $m_u=120$ and no shift begins at the tip of the arrow. Sequences 
with short-dashed lines correspond to composite models with the younger
population at $0$ (top), $3$ (middle) and $4.5$ (bottom) Myr age superposed on
an older population. The older population is 6--14~Myr old along the sequence.
Each sequence begins with a dot marked by vertical line.}
%
\figcaption{
Observed regions in $\log$(\philb)$_3$ vs $(V-R)_3$ plane. See Fig.~6 for 
other details. }
%
\figcaption{
Histogram of masses of GEHRs. }
%
\figcaption{
A schematic diagram of the four phases during star-formation in GEHRs. The
outer boundary in each sketch represents a Giant Molecular Association which
contains a GEHR, surrounded by the low density gas. The GEHR itself contains
cluster stars, clumpy gas and dust as shown. The formation of GEHR core, 
subsequent formation of the halo, the spreading out of optically
luminous stars into lower density regions, and subsequent triggering of a 
second burst of star formation in the left out gas are some of the salient
features of the model. Typical time scales involved are marked on the top 
right. See section 5.2 for details.}

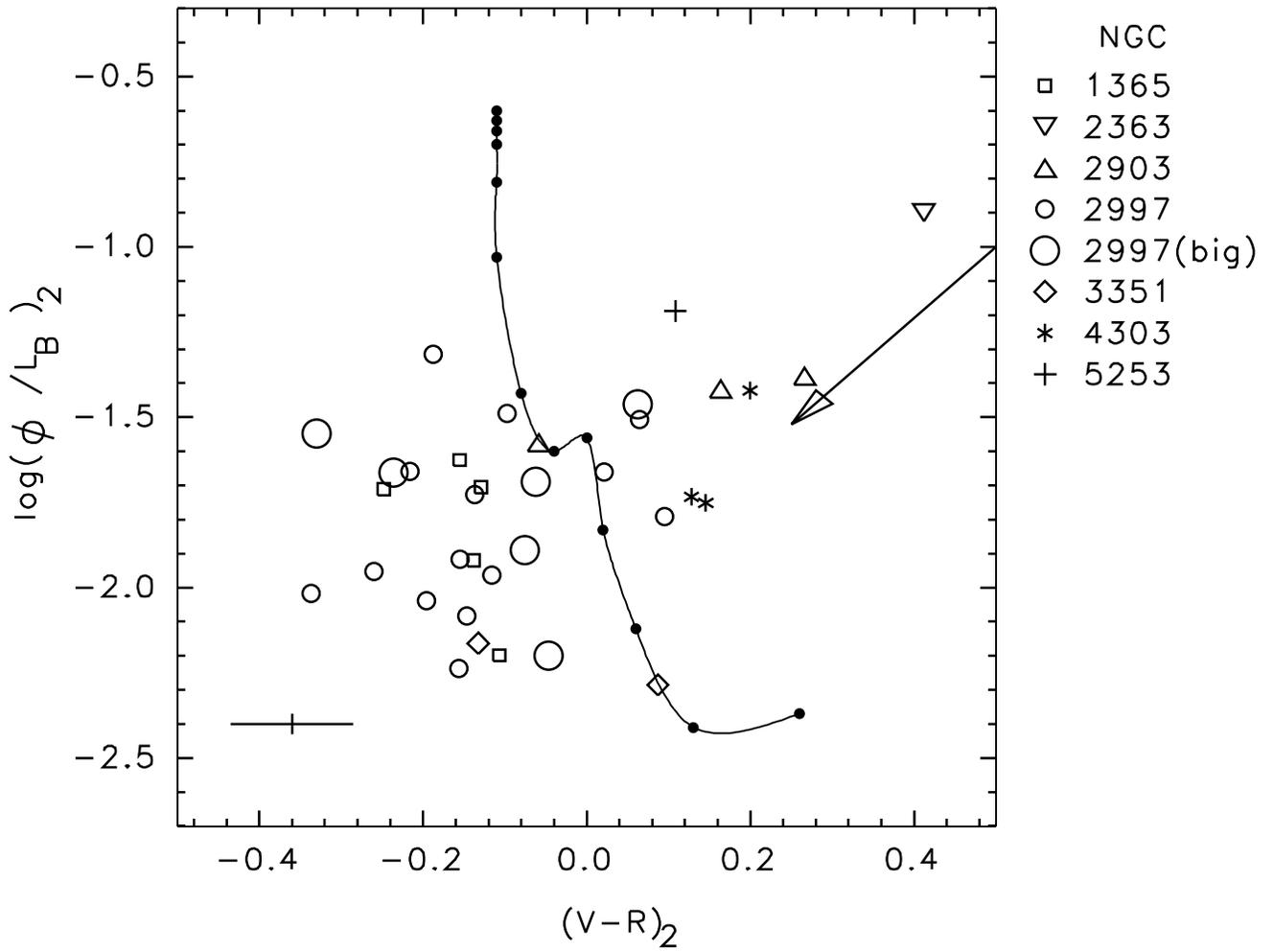
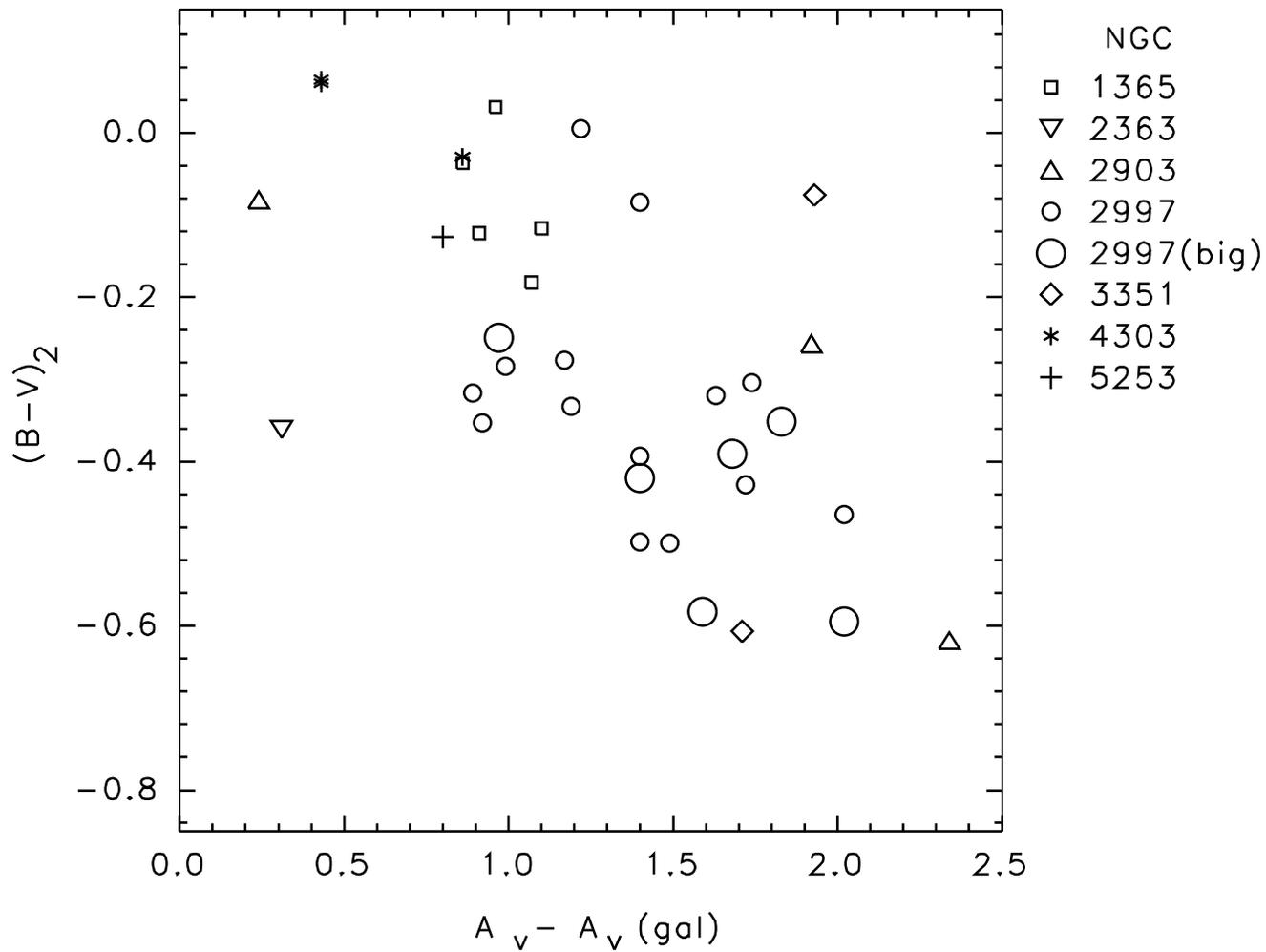
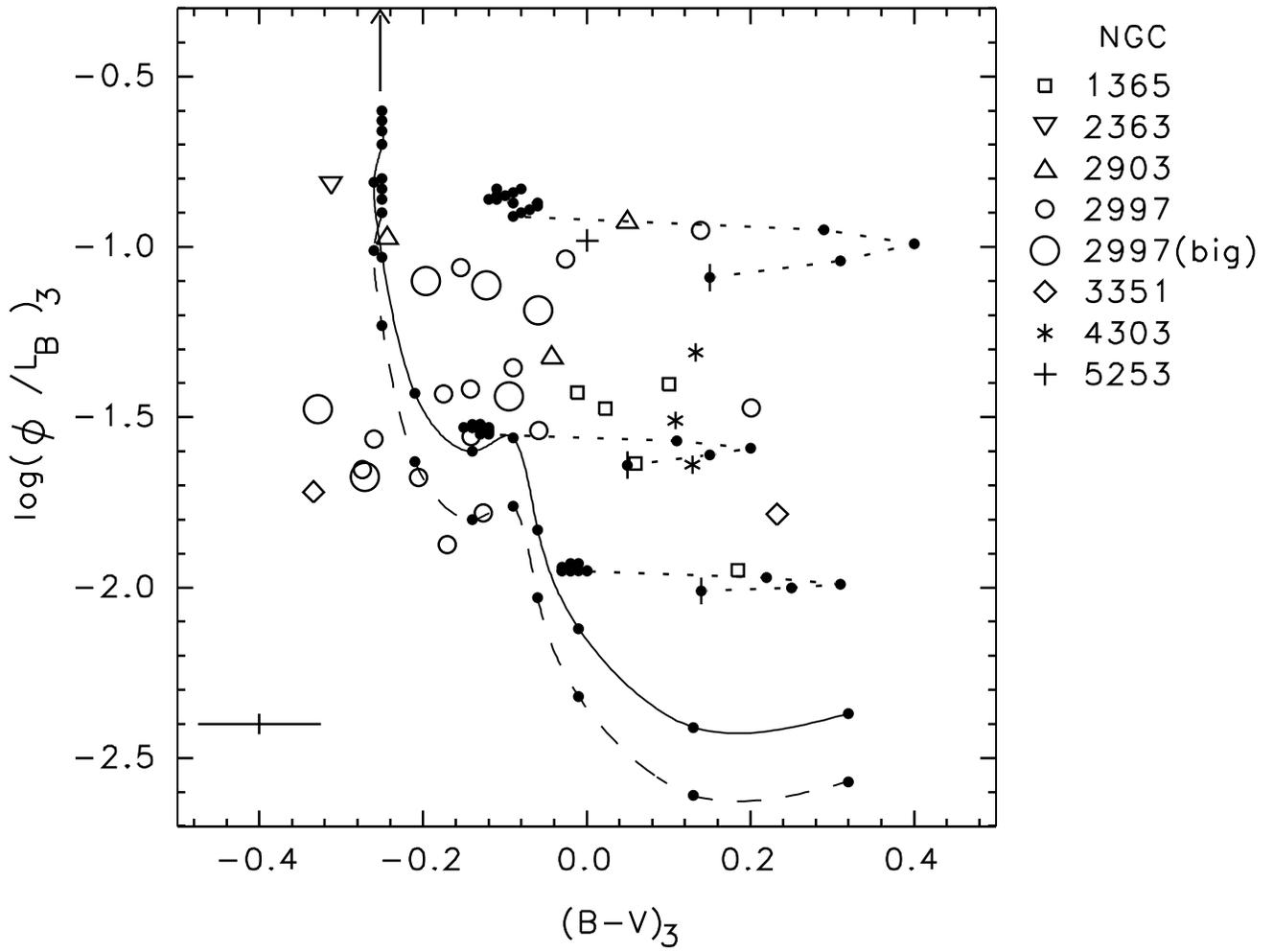
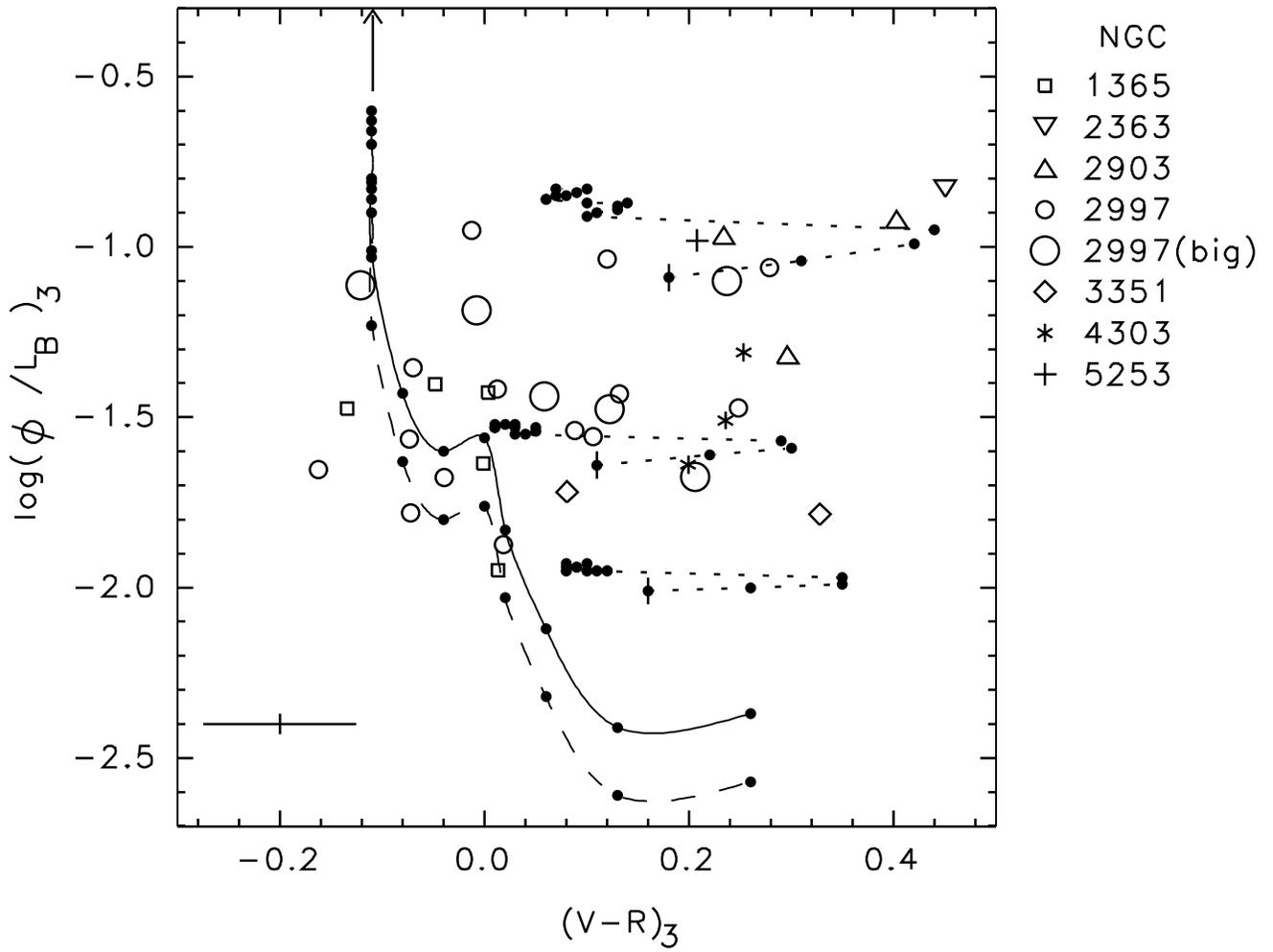
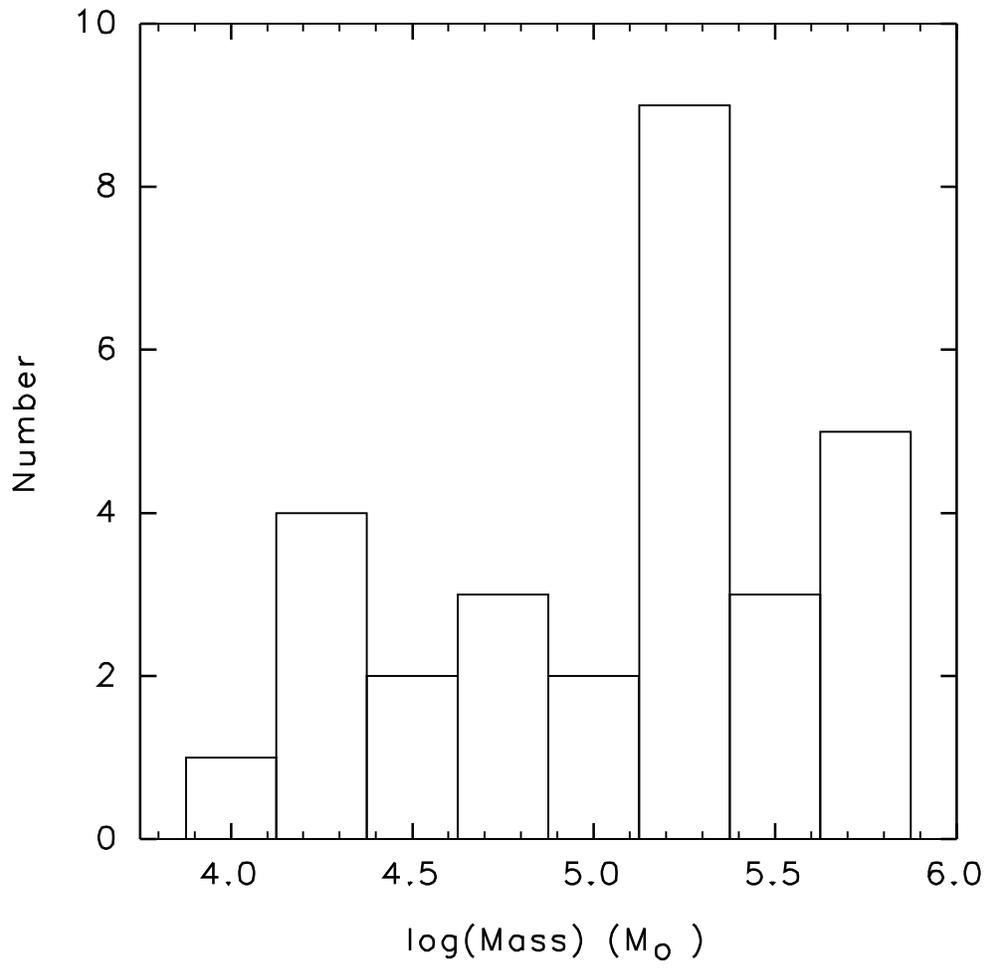
\end{document}